# OPPORTUNISTIC SCHEDULING AND BEAMFORMING FOR MIMO-SDMA DOWNLINK SYSTEMS WITH LINEAR COMBINING

Man-On Pun, Visa Koivunen and H. Vincent Poor

## Abstract

Opportunistic scheduling and beamforming schemes are proposed for multiuser MIMO-SDMA downlink systems with linear combining in this work. Signals received from all antennas of each mobile terminal (MT) are linearly combined to improve the *effective* signal-to-noise-interference ratios (SINRs). By exploiting limited feedback on the effective SINRs, the base station (BS) schedules simultaneous data transmission on multiple beams to the MTs with the largest effective SINRs. Utilizing the extreme value theory, we derive the asymptotic system throughputs and scaling laws for the proposed scheduling and beamforming schemes with different linear combining techniques. Computer simulations confirm that the proposed schemes can substantially improve the system throughput.

## I  Introduction

Multiple-input multiple-output (MIMO) technology employing multiple transmit and receive antennas has emerged as one of the most promising techniques for broadband data transmissions in wireless communication systems [1]. In particular, recent studies have shown that MIMO can substantially increase the sum-rate capacity of a downlink system where a base station (BS) communicates simultaneously with multiple mobile terminals (MTs) [2]. However, the capacity achieving strategy using dirty paper coding not only incurs high computational complexity but also requires perfect channel state information available to the BS [2]. To circumvent these obstacles, *opportunistic beamforming* with proportional fair scheduling (OB-PFS) has been proposed in [3] as an effective means of achieving the asymptotic sum-rate capacity by exploiting *multiuser diversity* with limited channel feedback. In OB-PFS, the downlink transmission time is divided in slots comprised of mini-slots. Users' channels are assumed to be approximately invariant during one slot but may vary from one slot to another. In the beginning of each slot, the BS broadcasts one pilot symbol weighted by a randomly generated complex vector (also referred to as the random beam). Then, each MT evaluates the signal-to-noise ratio (SNR) by exploiting the pilot signal and feeds back the SNR information to the BS. Taking into account fairness, the BS schedules data transmission to the MT with the best normalized instantaneous channel condition throughout the rest of the slot. Recently, some extensions of [3] employing *multiple* beams have been developed [4,5]. Regardless of the number of beams employed in [3–5], these opportunistic schemes schedule only one MT in each slot, and so can generally be considered to be the time-sharing scheduling schemes (TS-SS). In contrast, [6] has proposed an opportunistic space-division multiple access-based scheduling scheme (OSDMA-SS) employing multiple orthonormal beams to serve multiple MTs *simultaneously* in each slot. Denote by $M$ and $N$ the number of transmit and receive antennas, respectively. It has been shown recently that the sum-rate of OSDMA-SS grows linearly with $M$ whereas that of TS-SS increases only linearly with $\min(M, N)$ [7]. In addition to the more rapidly growing scaling law, OSDMA-SS is particularly attractive for practical systems with stringent latency requirements since multiple users can be served during each time slot.

In OSDMA-SS, the BS broadcasts pilot signals weighted by multiple orthonormal beams in the beginning of each time slot [6]. For each single-antenna MT, it evaluates the signal-to-interference-noise ratio (SINR) on each beam and feeds back information on its desired beam with the highest SINR. Assuming that each beam is requested by at least one MT, the BS awards each beam to the MT with the highest corresponding SINR among all MTs. For MTs with multiple receive antennas, [6] proposes to let each antenna compete for its desired beam as if it were an individual MT. As a result, each beam is assigned to a specific receive antenna of a chosen MT. Since signals received from the undesignated antennas of a chosen MT are discarded, [6] entails inefficient utilization of multiple receive antennas.

In this work, we propose opportunistic beamforming and scheduling schemes with different linear combining techniques for MIMO-SDMA downlink systems. In contrast with [6], signals received from all antennas of a chosen MT are jointly exploited to improve the effective SINR through the use of low-complexity linear combining techniques. Then, the $M$ effective SINRs are returned to the BS and employed as the scheduling metric. Using the extreme value theory, we prove that the cumulative distribution functions (CDFs) of the effective signal-to-interference ratios (SIRs) obtained with different linear combining techniques converge asymptotically to the Frechet-type limiting distributions. Based on the limiting distributions, we derive the asymptotic throughput and closed-form scaling laws for the proposed opportunistic beamforming and scheduling schemes.

*Notation*: Vectors and matrices are denoted by boldface letters. $\|\cdot\|$ represents the Euclidean norm of the enclosed vector. $\boldsymbol{I}_N$ is the $N \times N$ identity matrix. $[\boldsymbol{a}]_i$ indicates the $i$th entry of vector $\boldsymbol{a}$. We use $E\{\cdot\}$, $(\cdot)^T$ and $(\cdot)^H$ for expectation, transposition and Hermitian transposition. Finally, $\log$ and $\ln$ are the logarithms to the base 2 and $e$, respectively, and $|\cdot|$ denotes the amplitude of the enclosed complex-valued quantity.

Man-On Pun and H. Vincent Poor are with the department of Electrical Engineering, Princeton University, Princeton, NJ 08544.

Visa Koivunen is on sabbatical leave at Princeton University from Helsinki University of Technology (HUT), Finland.

This research was supported in part by the Croucher Foundation under a post-doctoral fellowship, and in part by the U. S. National Science Foundation under Grant No. ANI-03-38807.

## II  Signal Model

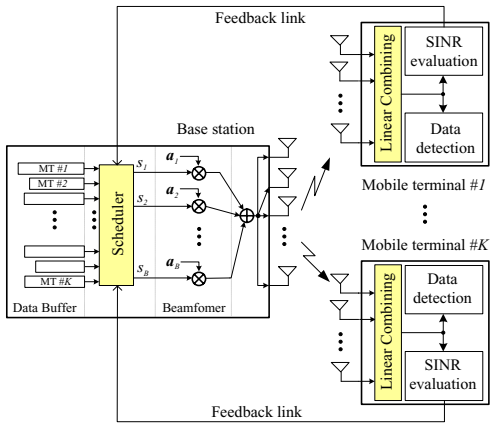

Figure 1: A block diagram of the opportunistic MIMO SDMA downlink system under consideration.

We consider the opportunistic MIMO-SDMA downlink system depicted in Fig. 1 where the BS is equipped with $M$ transmit antennas and each of the $K$ MTs has $N$ receive antennas with $N \leq M$. Let $\{\boldsymbol{a}_m; m = 1, 2, \cdots, M\}$ be a vector set containing $M$ orthornormal beamforming vectors of length $M$. We focus on a particular time slot during which a beamforming vector set $\{\boldsymbol{a}_m\}$ has been chosen from a common codebook shared by the BS and MTs. During the $p$th mini-slot, the transmitted signal can be expressed as

$$\boldsymbol{x}(p) = \sum_m \boldsymbol{a}_m s_m(p) = \boldsymbol{A}\boldsymbol{s}(p), \quad (1)$$

where $\boldsymbol{A} = [\boldsymbol{a}_1, \boldsymbol{a}_2, \cdots, \boldsymbol{a}_M]$ is the orthornormal beamforming matrix with $\boldsymbol{A}^H \boldsymbol{A} = \boldsymbol{I}_M$ and $\boldsymbol{s}(p) = [s_1(p), s_2(p), \cdots, s_M(p)]^T$ with $E\{|s_m(p)|^2\} = 1$ is the data vector transmitted in the $p$th mini-slot. Thus, the total transmission power is $M$.

For notational simplicity, we drop the temporal index $p$ in the sequel. Assuming that each MT experiences independent and identically-distributed (i.i.d.) frequency-flat Rayleigh fading, we use $\boldsymbol{h}_{k,j}$ to denote the $1 \times M$ channel gain vector with the $i$th entry, $[\boldsymbol{h}_{k,j}]_i$, representing the channel gain from the $i$th transmit antenna of the BS to the $j$th receive antenna of the $k$th MT. Thus, the signal received by the $j$th receive antenna of the $k$th MT takes the following form.

$$y_{k,j} = \boldsymbol{h}_{k,j}\boldsymbol{x} + n_{k,j}, \quad 1 \leq j \leq N. \quad (2)$$

where $n_{k,j}$ is zero-mean Gaussian noise with variance $\sigma_k^2$. $n_{k,j}$ is assumed to be statistically independent across receive antennas and users.

Collecting signals received from $N$ antennas into one vector, we can rewrite (2) into the following matrix form.

$$\boldsymbol{y}_k = \boldsymbol{H}_k \boldsymbol{x} + \boldsymbol{n}_k, \quad (3)$$

where $\boldsymbol{y}_k = [y_{k,1}, y_{k,2}, \cdots, y_{k,N}]^T$, $\boldsymbol{n}_k = [n_{k,1}, n_{k,2}, \cdots, n_{k,N}]^T$ and $\boldsymbol{H}_k$ is the $k$th MT's channel matrix defined as $\boldsymbol{H}_k = \left[\boldsymbol{h}_{k,1}^T, \boldsymbol{h}_{k,2}^T, \cdots, \boldsymbol{h}_{k,N}^T\right]^T$. In this work, we assume that the each MT has obtained perfect knowledge of its own channel matrix $\boldsymbol{H}_k$ by some means, e.g. training. Thus, the SINR of the $i$th beam perceived by the $j$th receive antenna of the $k$th MT is given as

$$\rho_{k,j,i} = \frac{|\boldsymbol{h}_{k,j}\boldsymbol{a}_i|^2}{\sum_{m \neq i} |\boldsymbol{h}_{k,j}\boldsymbol{a}_m|^2 + \sigma_k^2}, \quad (4)$$

for $k = 1, 2, \cdots, K$, $j = 1, 2, \cdots, N$ and $m = 1, 2, \cdots, M$. In the sequel, $\rho_{k,j,i}$ is referred to as the *measured* SINR whereas the *effective* SINR stands for the SINR obtained by linearly combining signals from all receive antennas. It will become evident in the following analysis that $\rho_{k,j,i}$ is less than the effective SINR. As a result, the scheduling scheme based on $\rho_{k,j,i}$ is suboptimal in terms of system throughput, compared to the proposed scheme using the effective SINR.

It is worthwhile to point out an interesting remark related to (3). Despite the similarity of (3) and the signal model commonly used in the conventional point-to-point MIMO systems [1], the subtle difference lies in that the channel matrix in (3) is fat (more columns than rows) whereas that in [1] is tall (more rows than columns). As a result, the MTs in the MIMO-SDMA systems have only $N$ degrees of freedom to suppress maximum $M - 1$ interfering beams. Without effective interference suppression, the system performance will be degraded significantly due to the lack of degrees of freedom for interference suppression. For systems with a large number of MTs, this problem can be alleviated by scheduling data transmissions to MTs whose channels are orthogonal. However, for systems with a few MTs, effective interference suppression becomes particularly crucial to achieving high system performance.

## III  Proposed Scheme

In this section, we develop a new beamforming and scheduling scheme for MIMO-SDMA by exploiting the effective SINR obtained with linear combining techniques. In the beginning of each time slot, each MT evaluates the effective SINR for each beam and feeds back information about $M$ effective SINRs to the BS. More specifically, the effective SINRs of the $i$th beam at the $k$th MT obtained with selection combining (SC), maximum ratio combining (MRC) and optimum combining (OC) techniques can be computed as follows [8]:

$$\gamma_{k,i}^{(\text{SC})} = \max_{1 \leq n \leq N} \left\{ \frac{|\boldsymbol{h}_{k,n}\boldsymbol{a}_i|^2}{\sum_{m \neq i} |\boldsymbol{h}_{k,n}\boldsymbol{a}_m|^2 + \sigma_k^2} \right\}, \quad (5)$$

$$\gamma_{k,i}^{(\text{MRC})} = \frac{\|\boldsymbol{H}_k\boldsymbol{a}_i\|^4}{\sum_{m \neq i} |\boldsymbol{a}_i^H \boldsymbol{H}_k^H \boldsymbol{H}_k \boldsymbol{a}_m|^2 + \|\boldsymbol{H}_k\boldsymbol{a}_i\|^2 \sigma_k^2}, \quad (6)$$

$$\gamma_{k,i}^{(\text{OC})} = \boldsymbol{a}_i^H \boldsymbol{H}_k^H \boldsymbol{R}_{k,i}^{-1} \boldsymbol{H}_k \boldsymbol{a}_i, \quad (7)$$

where

$$\boldsymbol{R}_{k,i} = \boldsymbol{H}_k \sum_{m \neq i} \boldsymbol{a}_m \boldsymbol{a}_m^H \boldsymbol{H}_k^H + \sigma_k^2 \boldsymbol{I}_N. \quad (8)$$

Note that (7) performs active interference suppression by exploiting the interference structure in (8), whereas (5) and (7) simply intend to amplify the desired signal. It will be shown later that this characteristic interference-suppression feature of OC enables the scheduling scheme with OC to considerably outperform those with SC and MRC.

Upon receiving the effective SINR information from all MTs, the BS schedules and starts data transmission to multiple MTs with the largest effective SINRs on different beams until the end of the current time slot. At each chosen MT, received signals from all antennas are linearly combined using one of the above linear combining techniques, followed by data detection. It is worth noting that the probability of awarding multiple beams to the same MT is rather small, as the number of MTs is large. Furthermore, recall that the minimum mean squared error (MMSE) and zero-forcing (ZF) receiver structures for MIMO receivers amount to combiners using OC and MRC for each beam, respectively [8]. As a result, for an MT assigned with multiple beams, it can focus on one assigned beam at a time using the chosen combining technique while regarding all other beams as interfering sources.

## IV Throughput Analysis

Define $\gamma_m^* = \max(\gamma_{1,m}, \gamma_{2,m}, \cdots, \gamma_{K,m})$, for $m = 1, 2, \cdots, M$. Assuming $\gamma_{k,m}$ for $k = 1, 2, \cdots, K$, are i.i.d. with CDF $F_X(x)$, the resulting average system throughput can be computed as [2]:

$$C = E\left\{\sum_{m=1}^{M} \log(1 + \gamma_m^*)\right\}, \qquad (9)$$

$$= M \int_0^\infty \log(1+x) \, d\left[F_X(x)\right]^K. \qquad (10)$$

In the following, we first derive the $F_X(x)$ functions obtained with different combining techniques before establishing their corresponding limiting distributions.

### A CDF analysis in interference-limited environment

To keep our analysis tractable, we consider the interference-limited scenario in which the interference power is much larger than the additive noise power, which generally holds for $M > N$. As a result, our following analysis concentrates on the effective SIR, rather than the effective SINR.

The CDFs of the effective SIRs obtained with different combining techniques can be derived as follows.

$$F_X^{(\text{Measured})}(x) = 1 - \frac{1}{(1+x)^{M-1}}, \qquad (11)$$

$$F_X^{(\text{SC})}(x) = \left[1 - \frac{1}{(1+x)^{M-1}}\right]^N, \qquad (12)$$

$$F_X^{(\text{MRC})}(x) = 1 - \frac{1}{(1+x)^{M-1}} - \sum_{p=1}^{N-1} \binom{M+N-p-2}{M-2} \frac{x^{N-p}}{(1+x)^{M+N-p-1}}, \qquad (13)$$

$$F_X^{(\text{OC})}(x) = 1 - \frac{1}{(1+x)^{M-N}} - \sum_{p=1}^{N-1} \binom{M-p-1}{M-N-1} \frac{x^{N-p}}{(1+x)^{M-p}}, \qquad (14)$$

where (11)-(13) hold for $M \geq N$ while (14) requires $M > N$.

Subsequently, the CDF of the maximum SIR obtained with different combining techniques, $[F_X(x)]^K$, can be directly computed from (12)-(14) correspondingly.

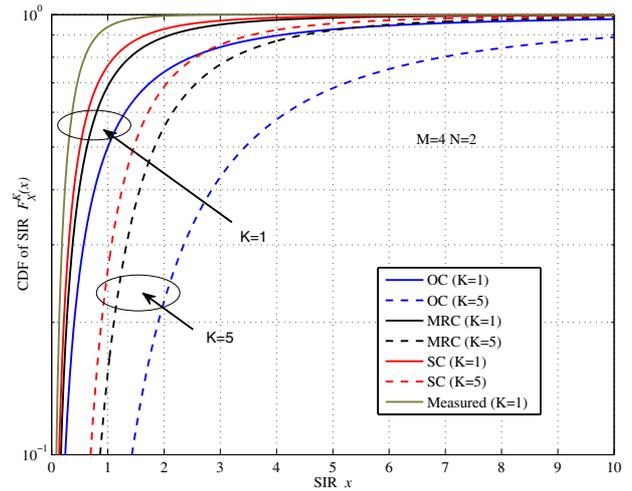

Figure 2: CDF's of SIR obtained with different combining techniques for $M = 4$ and $N = 2$.

Figure 2 shows the CDFs of the maximum effective SIR, $[F_X(x)]^K$, derived from (11)-(14) for the case of $M = 4$, $N = 2$ and $K = 1, 5$. The selection of $M = 4, N = 2$ is due to the considerable practical interests in systems with those parameters. Figure 2 indicates that the CDF of the effective SIR obtained with OC has a heavier tail than those obtained with MRC and SC. As a result, it is more probable for OC to achieve a larger effective SIR than the others for given $M$ and $N$, which leads to a higher system throughput. Furthermore, it is evident from Fig. 2 that the tail behavior of OC improves more significantly over MRC and SC as $K$ increases.

It is interesting to compare the proposed scheme with SC and that proposed in [6]. In [6], a receive antenna at each MT is regarded as an individual MT. Under some mild assumptions, it has been shown in [6] that the maximum SINR over all beams measured at a particular receive antenna is equal to the maximum SINR for that best beam measured at all $N$ receive antennas. As a result, the CDF of the maximum effective SINR for the $i$th beam at the $k$th MT can be approximated by

$$\gamma_{k,i}^{(\text{SH})} \approx \left[1 - \frac{1}{(1+x)^{M-1}}\right]^{NK_i}, \qquad (15)$$

where $K_i \leq K$ is the number of requests for the $i$th beam from the $K$ MTs.

Comparison between (12) and (15) reveals that [6] can be considered as a suboptimal form of the proposed scheme with selection combining. This suboptimality is caused by the fact that the SC-based scheduling scheme under consideration requires information on all $M$ beams whereas [6] simply feeds back information on $N$ beams. However, as will be shown shortly, even the performance of the SC-based scheduling scheme is rather unimpressive compared to schemes employing other combining techniques.

*B Asymptotic throughput*

We proceed to investigate the asymptotic behavior of $F_{X_{(K)}}(x) = [F_X(x)]^K$ as $K$ increases. To shed light on the performance of the proposed opportunistic scheduling schemes with tractable analytical complexity, we concentrate the following analysis on a system of high practical interest, *i.e.* $M = 4$ and $N = 2$. The results for other values of $M$ and $N$ can be obtained in a similar fashion. Due to space limitations, we will discuss the derivation for the case of OC and provide only the final results for the other two combining techniques.

For $M = 4$ and $N = 2$, (14) becomes

$$F_X^{(OC)}(x) = 1 - \frac{1+3x}{(1+x)^3}. \tag{16}$$

It is known in the context of extreme value theory that the limiting distribution of $F_{X_{(K)}}(x) = [F_X(x)]^K$, if it exists, is one of three types [9]. Fortunately, we can easily prove that the parent distribution given in (16) is of the Pareto type and satisfies the following equation

$$\lim_{x \to \infty} \frac{1 - F_X^{(OC)}(x)}{1 - F_X^{(OC)}(cx)} = c^2, \quad c > 0 \tag{17}$$

which is a necessary and sufficient condition for the resulting limiting distribution being of the Frechet type. Consequently, $F_{X_{(K)}}(x) = [F_X(x)]^K$ converges to the following Frechet-type distribution [9].

$$F_{X_{(K)}}^{(OC)}(a_K x) = \begin{cases} 0 & x \le 0 \\ \exp(-x^{-2}) & x > 0 \end{cases}, \tag{18}$$

where $a_K$ is a normalizing factor.

It is worth noting that the asymptotic analysis reported in [3] and [6] has been conducted in terms of the SINR, rather than the SIR. As shown in [3] and [6], the resulting limiting distribution in that case is of the Gumbel type since their corresponding parent distributions are of the exponential type.

The normalizing factor $a_K$ in (18) can be computed from the so-called characteristic extreme of (16) and is given by

$$F_X^{(OC)}(a_K) = 1 - \frac{1}{K}, \tag{19}$$
$$a_K^{(OC)} \approx \sqrt{3K} - 1. \tag{20}$$

Thus, we have

$$P_{X_{(K)}}^{(OC)}\left(X_{(K)} \le \left(\sqrt{3K} - 1\right)x\right) \approx e^{-x^{-2}}, \tag{21}$$

for $x \ge 0$, or equivalently,

$$P_{X_{(K)}}^{(OC)}(X_{(K)} \le x) \approx e^{-(\sqrt{3K}-1)^2 x^{-2}}. \tag{22}$$

Substituting (22) into (10), we have the average sum-rate obtained with OC as follows.

$$C^{(OC)} = 4 \int_0^\infty \log(1+x) \, d\left[e^{-\frac{(\sqrt{3K}-1)^2}{x^2}}\right], \tag{23}$$

$$= -4 \int_0^\infty \log(1+x) \, d\left[1 - e^{-\frac{(\sqrt{3K}-1)^2}{x^2}}\right], \tag{24}$$

$$= \frac{4}{\ln 2} \int_0^\infty \frac{1 - e^{-\frac{(\sqrt{3K}-1)^2}{x^2}}}{1+x} \, dx, \tag{25}$$

where the last equality is obtained by using the integration by parts. Since the closed-form expression in (25) is non-trivial, we evaluate the average sum-rate obtained with OC by resorting to numerical methods.

Similarly, we have the average sum-rates obtained with MRC and SC as follows.

$$C^{(MRC)} = \frac{4}{\ln 2} \int_0^\infty \frac{1 - e^{-\frac{\left(\sqrt[3]{4K}-1\right)^3}{x^3}}}{1+x} \, dx, \tag{26}$$

$$C^{(SC)} = \frac{4}{\ln 2} \int_0^\infty \frac{1 - e^{-\frac{\left(\sqrt[3]{2K}-1\right)^3}{x^3}}}{1+x} \, dx. \tag{27}$$

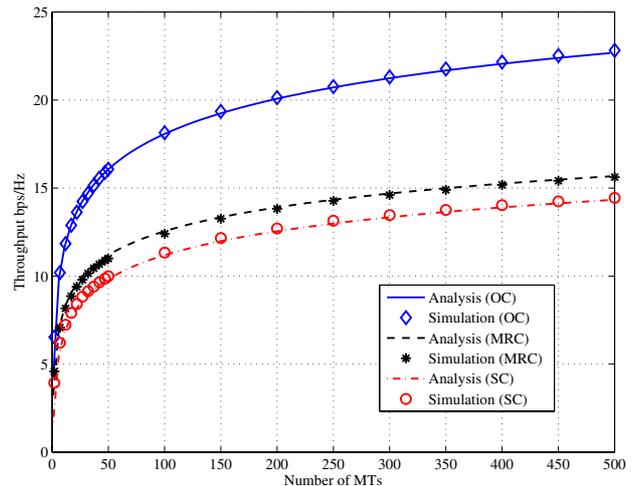

Figure 3: Comparison of analytical and simulated average throughput as a function of the number of MTs with $M = 4$, $N = 2$ and $\sigma_k^2 = 0$. The analytical and simulation results agree well with each other.

Figure 3 compares the average throughput obtained by analysis in (25)-(27) to the simulation results. It is evident from Fig. 3 that the analytical and simulated results are in accord with each other. Furthermore, inspection of Fig. 3 reveals that the throughput obtained with OC is near 50% larger than those obtained with SC and MRC for $M = 4$ and $N = 2$ in the noise-free scenario, which is attributed to the interference suppression capability built in OC.

*C Sum-rate scaling law*

To provide insights into the influence of the number of MTs, $K$, on the system throughput, we investigate the scaling laws of the proposed schemes, assuming $K$ is sufficiently large. Similar to the previous Section, we concentrate on the scaling law for the proposed scheme with OC whereas only the final results for those with SC and MRC are provided. To derive the scaling law of the scheme with OC, we first rewrite (25) by letting $z = \frac{1}{x}$, where $x \in (0^+, \infty)$. Thus, we have

$$C^{(\text{OC})} \approx \frac{4}{\ln 2} \int_{0^+}^{\infty} \frac{1 - e^{-\left(\sqrt{3K}-1\right)^2 z^2}}{(1+z)\,z} \, dz, \quad (28)$$

$$\approx \frac{4}{\ln 2} \int_{0^+}^{\frac{2}{\sqrt{3K}-1}} \frac{1 - e^{-\left(\sqrt{3K}-1\right)^2 z^2}}{(1+z)\,z} \, dx +$$

$$\frac{4}{\ln 2} \int_{\frac{2}{\sqrt{3K}-1}}^{\infty} \frac{1}{(1+z)\,z} \, dx, \quad (29)$$

where the last approximation is obtained by exploiting the fact that $1 - e^{-\xi} \approx 1$, for $\xi \geq 4$.

Taking the limit of (29) as $K$ tends to infinity, it is easy to show that the limit of the second term on the right hand side (R.H.S.) takes the following form

$$\lim_{K \to \infty} \frac{\frac{4}{\ln 2} \int_{\frac{2}{\sqrt{3K}-1}}^{\infty} \frac{1}{(1+z)z} \, dx}{4 \log\left(\sqrt{3K}\right)} = 1, \quad (30)$$

whereas the limit of the first term becomes negligibly small as $\lim_{K \to \infty} \frac{2}{\sqrt{3K}-1} = 0$. As a result, the scaling law for the proposed scheme with OC is given as

$$\lim_{K \to \infty} \frac{C^{(\text{OC})}}{4 \log\left(\sqrt{3K}\right)} = 1. \quad (31)$$

Similarly, we can obtain the scaling laws for the schemes with SC and MRC as follows.

$$\lim_{K \to \infty} \frac{C^{(\text{MRC})}}{4 \log\left(\sqrt[3]{4K}\right)} = 1, \quad (32)$$

$$\lim_{K \to \infty} \frac{C^{(\text{SC})}}{4 \log\left(\sqrt[3]{2K}\right)} = 1. \quad (33)$$

Figure 4 depicts the scaling laws obtained in (31), (32) and (33). Comparison between Figs. 3 and 4 indicates that the scaling laws are approximately on par with the simulation results.

### D  Fairness in scheduling

In this discussions above, we consider a homogeneous network where the channel gains for all MTs are i.i.d. Consequently, fair scheduling among all MTs is guaranteed. However, even for practical systems experiencing the near-far effect, the fairness of the proposed scheduling scheme can be shown by employing an approach similar to that in [6]. Intuitively speaking, since the system under consideration is interference-limited, the MT that is closer to the BS will receive stronger interference. Consequently, the effective SIR of an MT is mainly determined by the *alignment* of the random beams and its instantaneous channel matrix, rather than solely by the determinant of the channel matrix.

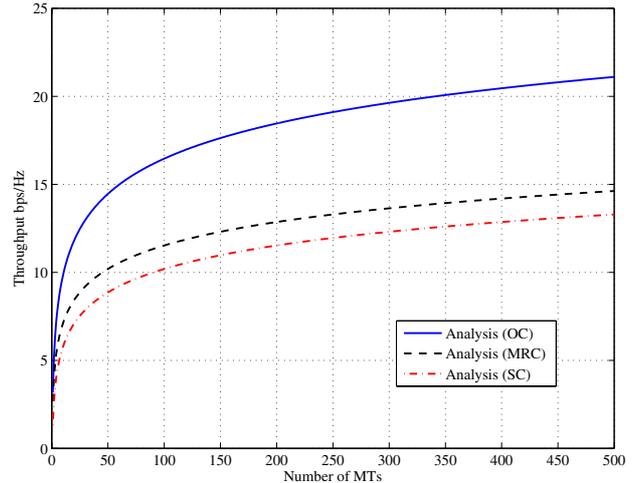

Figure 4: Scaling laws as a function of the number of MTs with $M = 4$ and $N = 2$.

## V  SIMULATION RESULTS

In this section, we use computer simulation to confirm the performance of the proposed scheduling and beamforming schemes in noisy environments. Unless otherwise specified, the noise variance is set to $\sigma_k^2 = 1$ for $k = 1, 2, \cdots, K$.

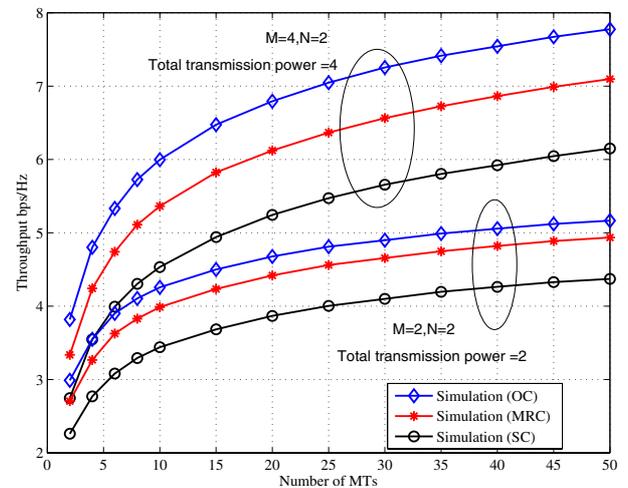

Figure 5: Average throughput as a function of the number of MTs with $M = 2, 4$, $N = 2$ and $\sigma_k^2 = 1$.

Figure 5 shows the average throughput of the proposed schemes with $M = 2, 4$ and $N = 2$. Fig. 5 indicates that the throughput of OC represents an impressive 20% and 10% increase compared to that of SC and MRC, respectively, at $K = 50$. Note that the total transmission power for $M = 4$ is twice of that for $M = 2$ in Fig. 5. As a result, the average SNR for each beam is approximately constant.

In contrast, Figure 6 depicts the average throughput of the proposed schemes with $M = 2, 4$ and $N = 2$ for a *fixed* to-tal transmission power of 2. It is evident from Fig. 6 that the throughput with $M = 4$ is higher than that with $M = 2$ for the same amount of total transmission power. This is because the effect of an increasing $M$ on the average throughput is twofold. On the one hand, a larger $M$ creates more interference to a specific desired beam, which incurs loss of SINR. On the other hand, a larger $M$ also linearly increases the system throughput as indicated in (10), which outweighs the throughput loss due to a reduced SINR as shown in Fig. 6.

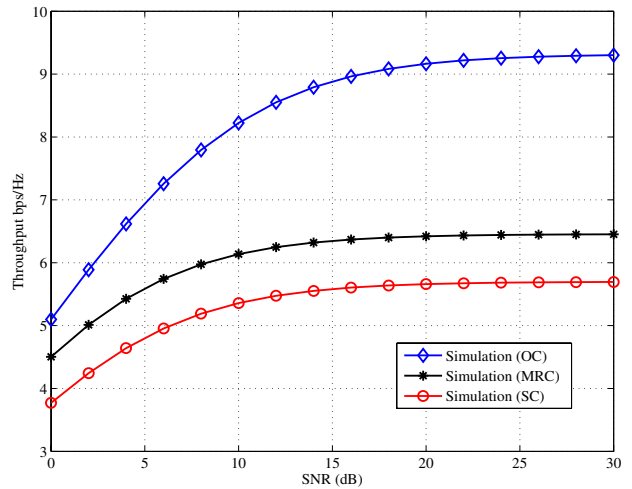

Figure 7: Average throughput as a function of SNR with $M = 4$, $N = 2$ and $K = 5$.

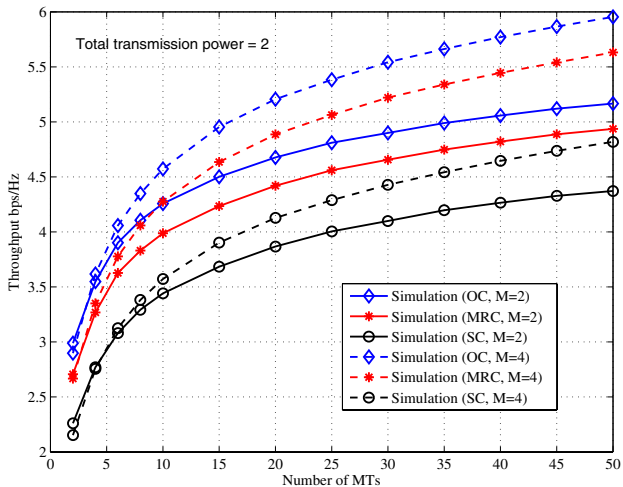

Figure 6: Average throughput as a function of the number of MTs with $M = 2, 4$, $N = 2$, $\sigma_k^2 = 1$ and a fixed total transmission power of 2.

In the last experiment, we consider a system with $M = 4$ and $N = 2$, serving a smaller number of MTs, i.e. $K = 5$. Figure 7 shows the average throughput of the proposed scheduling schemes as functions of SNR defined as $1/\sigma_k^2$. We can observe a substantial throughput gain provided by OC compared to SC and MRC. This is because the small number of MTs reduces the probability that several users' channel vectors are perfectly aligned to different orthogonal beams. As a result, the existence of stronger interference entails substantial SINR loss. Then, a receiver employing OC can provide more effective interference suppression, which results in considerable performance improvement over those with SC and MRC.

## VI CONCLUSION

Opportunistic scheduling and beamforming schemes for MIMO-SDMA downlink systems with linear combining have been proposed in this work. Using the extreme value theory, we have shown that the limiting distribution of the effective SIRs obtained with linear combining is of the Frechet type. Furthermore, the system throughput and scaling laws for the proposed schemes are derived. In particular, for practical systems with $M = 4$ and $N = 2$, it has been shown that the throughput of the proposed scheme with OC scales like $4\log\left(\sqrt{3K}\right)$ whereas those with MRC and SC are governed by $4\log\left(\sqrt[3]{4K}\right)$ and $4\log\left(\sqrt[3]{2K}\right)$, respectively. It has been demonstrated through asymptotic analysis and computer simulation that incorporating low-complexity linear combining techniques into the design of scheduling schemes for MIMO-SDMA downlink systems can substantially increase the system throughput.